# Effects of transformed Hamiltonians on Hamilton-Jacobi theory in view of a stronger connection to wave mechanics

Michele Marrocco*

*ENEA*

*(Italian National Agency for New Technologies, Energies and Sustainable Economic Development)*

*via Anguillarese 301, I-00123 Rome, Italy*

&

*Dipartimento di Fisica, Università di Roma 'La Sapienza'*

*P.le Aldo Moro 5, I-00185 Rome, Italy*

**Abstract**

Hamilton-Jacobi theory is a fundamental subject of classical mechanics and has also an important role in the development of quantum mechanics. Its conceptual framework results from the advantages of transformation theory and, for this reason, relates to the features of the generating function of canonical transformations connecting a specific Hamiltonian $H$ to a new Hamiltonian $K$ chosen to simplify Hamilton's equations of motion. Usually, the choice is between $K = 0$ and a cyclic $K$ depending on the new conjugate momenta only. Here, we investigate more general representations of the new Hamiltonian. Furthermore, it is pointed out that the two common alternatives of $K = 0$ and a cyclic $K$ should be distinguished in more detail. An attempt is made to clearly discern the two regimes for the harmonic oscillator and, not surprisingly, some correspondences to the quantum harmonic oscillator appear. Thanks to this preparatory work, generalized coordinates and momenta associated with the Hamilton's principal function $S$ (i.e., $K = 0$) are used to determine dependences in the Hamilton's characteristic function $W$ (i.e., cyclic $K$). The procedure leads to the Schrödinger's equation where the Hamilton's characteristic function plays the role of the Schrödinger's wave function, whereas the first integral invariant of Poincaré takes the place of the reduced Planck constant. This finding advances the pedagogical value of the Hamilton-Jacobi theory in view of an introduction to Schrödinger's wave mechanics without the tools of quantum mechanics.

## I. INTRODUCTION

Hamilton-Jacobi theory is very well known for being one of the main subjects of classical mechanics. Its popularity is largely due to an equation, the so-called Hamilton-Jacobi equation, that condenses many advantages of the theory of canonical transformations into a concise mathematical description of mechanical problems. The basics of this approach are available in many textbooks and synopses used to teach advanced classical mechanics. To name some relevant sources, it is common to refer to the popular textbook by Goldstein and coauthors[1] and to the volume dedicated to mechanics in the Landau & Lifshitz series[2] as the main resource books for students. Excellent material, prepared for specific courses, is also available on-line.[3] Not to name other textbooks that present the Hamilton-Jacobi approach with sufficient depth.[4-8]

Despite the large availability of information about the subject, a brief description of Hamilton-Jacobi theory is given here to ease the understanding of the argument put forward in this work. Readers who are familiar with the conceptual elements leading to the Hamilton-Jacobi equation can actually skip this introduction.

The starting point is represented by the canonical transformations that are summarized below according to the notation of Ref. 1. They are useful when the Hamilton's equations of motion

$$\begin{cases} \dot{q}_i = \dfrac{\partial H}{\partial p_i} \\ \dot{p}_i = -\dfrac{\partial H}{\partial q_i} \end{cases} \qquad i = 1,2,\dots,n \qquad (1)$$

are difficult to solve for the given Hamiltonian $H = H(q,p,t)$ (with $q = \{q_i\}$ and $p = \{p_i\}$). If so, the $2n$ equations in Eq. (1) might be transformed into a new set of Hamilton's equations

$$\begin{cases} \dot{Q}_i = \dfrac{\partial K}{\partial P_i} \\ \dot{P}_i = -\dfrac{\partial K}{\partial Q_i} \end{cases} \qquad (2)$$

whose solution is easier to obtain in comparison with Eq. (1). Of course, this approach works if we are able to find the transformation

$$\begin{cases} q_i = q_i(Q,P,t) \\ p_i = p_i(Q,P,t) \end{cases} \tag{3}$$

that connects the new variables $Q = \{Q_i\}$ and $P = \{P_i\}$ to the old ones. The transformed Hamiltonian $K$ is instead related to the old Hamiltonian $H$ by means of the following equation

$$p_i \dot{q}_i - H = P_i \dot{Q}_i - K + \frac{dF}{dt} \tag{4}$$

that results from variational principles (i.e., modified Hamilton's principle). To appreciate Eq. (4), some explaining is necessary. First of all, summation is tacitly implied over the repeated indices. Secondly, we have disregarded scale transformations that imply a trivial rescaling of coordinates and momenta. More importantly for our argument, the letter $F$ represents the so-called generating function of the transformation and, for this reason, plays the fundamental role in the search for the relationships summarized in Eq. (3). In particular, $F$ is a function of any mixture of phase space variables belonging to the old and new sets of variables.

Of particular interest for what follows is the generating function of the second kind $F_2(q,P,t)$ that combines dependences on the old coordinates $q$ and new momenta $P$ through the relation $F = F_2(q,P,t) - Q_i P_i$. The interest in $F_2$ is manifest as soon as, after some algebraic manipulation, old momenta $p$ and new coordinates $Q$ are calculated according to

$$\begin{cases} p_i = \frac{\partial F_2}{\partial q_i} \\ Q_i = \frac{\partial F_2}{\partial P_i} \end{cases} \tag{5}$$

Therefore, it is understood that, if we gain information on the functional structure of $F_2$, Eq. (5) can be solved for $q$ and $p$ to get the solution reported in Eq. (3). One elegant way to fulfill this program revolves around the connection between old and new Hamiltonians through the partial time derivative of $F_2$

$$K = H + \frac{\partial F_2}{\partial t} \tag{6}$$

This equation follows from Eq. (4) and introduces the Hamilton-Jacobi equation that results from setting $K = 0$ and replacing the old momenta $p$ with the corresponding derivatives of $F_2$ appearing in Eq. (5). But, to complete our short summary, we conform to the broadly accepted notation that employs the capital letter $S$ to indicate the solution of Eq. (6), namely $F_2$. Thus, $S$ defines the so-called Hamilton's principal function and, in the end, the Hamilton-Jacobi equation reads

$$H\left(q, \frac{\partial S}{\partial q}, t\right) + \frac{\partial S}{\partial t} = 0 \tag{7}$$

This is a non-linear partial differential equation of the first order whose solution $S$ seems to be a function of the old coordinates $q$ and time only. Remarkably, the other variables (new momenta $P$) of $S$ do not appear in Eq. (7). However, we expect to find a method to identify them (or some functions of them).

The choice of equating the new Hamiltonian $K$ to zero has some other consequences beyond Eq. (7). For instance, it determines a nice and significant result for the total time derivative of the Hamilton's principal function

$$\frac{dS}{dt} = p_i \dot{q}_i - H = L \tag{8}$$

with $L$ the Lagrangian associated with the original Hamiltonian $H$. Moreover, for time-independent Hamiltonians $H(q,p) = E$, it is found that

$$S(q, P, t) = W(q, P) - Et \tag{9}$$

where the Hamilton's characteristic function

$$W(q, P) = \int p_i \, dq_i \tag{10}$$

is introduced. Based on Eq. (9), the partial derivative of $S$ with respect to time can be extracted

$$\frac{\partial S}{\partial t} = -E \tag{11}$$

so that the Hamilton-Jacobi equation can be rewritten in its restricted version

$$H\left(q, \frac{\partial W}{\partial q}\right) = E \tag{12}$$

For the sake of our argument, it is important to underline that Eqs. (7)-(11) are all consequences of the arbitrary choice established by letting the new Hamiltonian $K$ be zero. The choice is so peculiar that, in this work, we first examine some concerns emerging from this customary approach to Hamilton-Jacobi theory (Section II). We might also ask the question whether we expect some changes in the Hamilton-Jacobi theory if we make use of other conditions on the transformed Hamiltonian of time-dependent problems (Section III). For time-independent problems, we try instead to formulate the Hamilton-Jacobi method without the initial distinction between $K = 0$ and $K \neq 0$. This attempt is conceived for the purpose of an unitary presentation of the method (Section IV). The special features intrinsic to the choices of $K = 0$ and $K \neq 0$ are finally analyzed with reference to the Hamilton's functions $S$ (i.e., $K = 0$) and $W$ (i.e., cyclic $K$) of the harmonic oscillator (Section V). These special features emerging in both Hamilton's functions point to a direct connection with quantum mechanics. For this reason, the second part of this work is aimed at defining this connection. The conceptual tools, based on the distinction between $S$ and $W$, are first developed for the free particle and an equation is obtained where the Hamilton's characteristic function $W$ emulates the Schrödinger's wave function, whereas the first integral invariant of Poincaré replaces the reduced Planck constant (Section VI). The same procedure is applied to the harmonic oscillator (Section VII) and finally extended to broader physical contexts (Section VIII).

## II. QUESTIONS REGARDING THE TRANSFORMED HAMILTONIAN

The conventional Hamilton-Jacobi approach summarized above might raise doubts about the general value of the choice $K = 0$. First of all, with reference to a generic time-dependent Hamiltonian $H(q, p, t)$, the constraint that the transformed Hamiltonian $K$ is identically zero clashes apparently with the physical meaning of a mechanical system whose energy $H$ changes in time in one (inertial) reference frame and, conversely, is constantly zero in another reference frame. In other terms, if we have a non-conservative system and are able to find a canonical transformation so that the transformed energy of the system cancels out, then the new coordinates and momenta might not be all constants of the motion in remarkable contradiction with Eq. (2). On the other hand, if it were possible, it would appear that the nature of the system has changed by virtue of the canonical transformation. This scenario sounds strange. Before the transformation, the physical system undergoes loss and/or gain of energy in direct and explicit dependence on the time variable. After the canonical transformation, we have a conservative system whose Hamiltonian is now constant in time, although equal to zero. This sort of change seems so peculiar.

Strange as it may seem, let us examine one conventional program laid out by some authors[1,3,4] to justify the introduction (and the powerful beauty) of the Hamilton-Jacobi method. It is suggested that the new constant variables $Q$ and $P$ satisfying Eq. (2) for $K = 0$ might be the initial values $q(0)$ and $p(0)$ of the old coordinates and momenta so that Eq. (3), with the substitution $Q = q(0)$ and $P = p(0)$, provides the actual solution to the motion. Despite the undoubtedly attractive perspective, the suggestion is prone to arbitrariness. As a matter of fact, we have to solve Eq. (7) and its general solution is (see the footnote on page 148 of Ref. 2)

$$S(q, P, t) = S_0(q, P, t) + f(P) \tag{13}$$

where $S_0$ is a particular solution to Eq. (7) and $f(P)$ can be any function of the conjugate momenta $P$. Note that Eq. (13) is plausible because the Hamilton-Jacobi equation in Eq. (7) depends only on the partial derivatives of $S$ with respect to the coordinates $q$ and time. For this reason, any additive function of the new momenta $P$ disappears in Eq. (7). This means that, for example, the $i$-th coordinate $Q_i$ or its equivalent $q_i(0)$, specified by the following equation

$$q_i(0) = Q_i = \left.\frac{\partial S}{\partial P_i}\right|_{t=0} = \left.\frac{\partial S_0}{\partial P_i}\right|_{t=0} + \frac{\partial f}{\partial P_i} \tag{14}$$

remains undetermined because of the arbitrariness of $f(P)$. More drastically, the derivative of $f$ could be non-invertible implying the impossibility of completing successfully the Hamilton-Jacobi plan that has in Eq. (3) its ultimate goal. Note that Landau and Lifshitz suggest to take the $n$ conditions $\partial S/\partial P_i = 0$ to prove that the indetermination of the general solution can be suppressed

(see again the footnote on page 148 of Ref. 2). Regrettably, this suggestion is impractical. The $n$ conditions $\partial S/\partial P_i = 0$ prevent the mechanical problem from being solved. In effect, the inversion of the canonical transformation mediated by the Hamilton's principal function $S$ is possible only if the Hessian matrix of $S$ yields a non-vanishing determinant (Hessian condition), that is $|\partial S^2/\partial q_j \partial P_i| \neq 0$.[5] In other terms, if the suggestion $\partial S/\partial P_i = 0$ was acceptable, then the Hessian condition would not be obeyed.

Even though we neglected the problem mentioned above, we would encounter another impediment in the application of the Hamilton-Jacobi method based on the equivalence between the new variables and the initial conditions applied to the old variables. With reference to the more common time-independent problems, the suggestion (even in a loose hypothetical sense!) of the identities $Q = q(0)$ and $P = p(0)$ to exemplify the usefulness of the method is totally incorrect. This can be explained in a very simple fashion. Within the time-independent context, it is customary to take a cyclic transformed Hamiltonian $K(P)$ (see for instance Ref. 1) and solve the restricted Hamilton-Jacobi equation of Eq. (12). But, it is easy to verify that the Hamilton's equations for the transformed Hamiltonian generate new coordinates $Q$ that vary linearly in time. Clearly, this fact contradicts the assumption of $Q = q(0) = constant$. The contradiction is so remarkable that the absence of a comment in the traditional literature is incredibly surprising. To the best of our knowledge, an exception is for the book of Sussman and Wisdom (see page 413 of Ref. 6). Therefore, for those cases where the Hamilton-Jacobi method is actually applied (i.e., physical systems of constant energy), the simplest choice of $Q = q(0)$ and $P = p(0)$ cannot be made.

Another concern appears when we deepen the meaning of the choice $K \neq 0$ made for time-independent problems. These are characterized by the constant value $E$ of the energy, that is

$$H(q,p) = E \qquad (15)$$

In this circumstance, the conventional approach assumes that the transformed Hamiltonian $K$ is again equal to $E$. This option has become a commonplace in Hamilton-Jacobi theory and it is a cornerstone that is firmly stated in several texts and reviews (for example see Ref. 1: "the new and old Hamiltonians are equal"). This assumption is not without consequences for the physical meaning of the Hamilton-Jacobi theory. It suffices to compare Eq. (11) with Eq. (6). The energy conservation $K = H = E \neq 0$ applied to the general result of Eq. (6) means that

$$\frac{\partial S}{\partial t} = 0 \tag{16}$$

which is in clear contrast with Eq. (11). Not to forget Eq. (4) that becomes

$$\frac{dS}{dt} = p_i \dot{q}_i + Q_i \dot{P}_i = p_i \dot{q}_i \tag{17}$$

and replaces Eq. (8) where it was instead suggested the equality between the total time derivative of the Hamilton's principal function and the Lagrangian. For this reason, the choices of $K = 0$ and $K = E$ should be clearly distinguished. In other terms, the introduction of the Hamilton's principal function $S$ is well-founded only when the choice $K = 0$ is made, otherwise the other option $K = E$ can only be referred to the use of the Hamilton's characteristic function $W$. Of course, the two Hamilton's functions are related but it is fundamental to keep the distinction between them (especially in the comparison between the different dependences on the corresponding $Q$ coordinates).

None of the reported concerns affects the validity of the Hamilton-Jacobi equation that remains a tool of great value. To show this in view of a strong connection with quantum mechanics, we first make an attempt to briefly revise Hamilton-Jacobi theory considering possible changes to address the concerns discussed above. The objective is to seek simple ways to generalize Hamilton-Jacobi theory without alteration to the consistent structure that supports the Hamilton-Jacobi equation whose solution will be later examined for different examples of physical problems to disclose a new connection to Schrödinger's wave mechanics.

## III. EFFECT OF TIME-DEPENDENT TRANSFORMED HAMILTONIANS ON HAMILTON-JACOBI EQUATION

For time-dependent Hamiltonian $H(q,p,t)$, a less stringent condition than $K = 0$ is represented by an explicit time dependence in the transformed Hamiltonian

$$K = K(P,t) \tag{18}$$

The idea is taken from Arnold[7] who hints at the possibility without facing the difficulties inherent in the explicit time dependence of $K$ (see page 260 in Ref. 7). The original suggestion is made in view of a Hamilton's principal function that depends on $q$ and $Q$ coordinates. Here, we have adapted that suggestion to the more usual argument of $S = S(q,P,t)$. Note, however, that in Eq. (18) the further dependence on conjugate momenta $P$ is useful to eliminate ambiguity in the definition of coordinates $Q$. Given this premise, the Hamilton-Jacobi equation is modified as follows

$$H\left(q, \frac{\partial S}{\partial q}, t\right) + \frac{\partial S}{\partial t} = K(P,t) \tag{19}$$

Apparently, the right-hand side of Eq. (19) complicates the search for a solution. We can circumvent the difficulty under the hypothesis that $K(P,t)$ is a regular function. If so, we can set

$$K(P,t) = \frac{\partial G(P,t)}{\partial t} = \frac{dG(P,t)}{dt} \tag{20}$$

where, in the last passage, replacement of the partial time derivative with the total time derivative is guaranteed by the constant values of the new momenta ($K$ is cyclic in the new coordinates $Q$). At this point, we can define a secondary Hamilton's principal function

$$\Sigma(q,P,t) = S(q,P,t) - G(P,t) \tag{21}$$

where $S$ solves Eq. (7). The secondary Hamilton's principal function $\Sigma$ is useful to duplicate the Hamilton-Jacobi equation resulting from the use of Eq. (21) in Eq. (19)

$$H\left(q, \frac{\partial \Sigma}{\partial q}, t\right) + \frac{\partial \Sigma}{\partial t} = 0 \tag{22}$$

with the advantage that, based on Hamilton's equation for the coordinates $Q$, we find

$$Q_i = \frac{\partial G(P,t)}{\partial P_i} \tag{23}$$

so that the indetermination underlined in the previous section disappears here.

Eq. (22) can now be solved according to the usual procedure used for the original Hamilton-Jacobi equation of Eq. (7). However, the solution must obey the further condition

$$\frac{\partial \Sigma(q,P,t)}{\partial P_i} = \frac{\partial S(q,P,t)}{\partial P_i} - \frac{\partial G(P,t)}{\partial P_i} = 0 \tag{24}$$

because the derivatives of $S$ and $G$ are both equal to $Q_i$. It is worthwhile to point out that, thanks to the independence of the function $G$ on the coordinates $q$, the dynamical information necessary to solve the equations of motion is unaltered and, indeed, we can easily relate the momenta $p$ to the derivative of $S$

$$p = \frac{\partial \Sigma}{\partial q} = \frac{\partial S}{\partial q} \tag{25}$$

The conclusion of this brief analysis is that the Hamilton-Jacobi method can be formally confirmed if the condition $K = 0$ of the original theory is relaxed in favor of a more general condition of a time-dependent transformed Hamiltonian.

A final remark that is germane to the current discussion regards the speculative character of the Hamilton-Jacobi theory for time-dependent Hamiltonians. Despite the general value of the theoretical considerations made when the time variable appears explicitly in the Hamiltonian, the practical importance of the Hamilton-Jacobi equation comes from its application to time-independent problems that are treated in the next section.

## IV. EFFECT OF TIME-INDEPENDENT TRANSFORMED HAMILTONIANS ON HAMILTON-JACOBI EQUATION

For a time-independent Hamiltonian $H(q,p)$, the conventional approach revolves around the search for a transformed Hamiltonian $K$ cyclic in all the coordinates.[1-8] As noted above, this choice has some consequences for the Hamilton-Jacobi theory. Indeed, according to Eq. (6), the identity $K = H$ implies that the partial time derivative of the Hamilton's principal function is zero. Instead, in remarkable contrast with the general value of Eq. (6), it is commonly taken Eq. (11) to hold even if attention is given to the choice $K = H$. In other words, this choice implies Eq. (16) and, based on such an equation, the two Hamilton's functions coincide, or $S = W$. Vice versa, the Hamilton's principal function is $S = W - Et$ when $K = 0$. The two alternatives for the transformed Hamiltonian $K$ cause different representations for the Hamilton's principal function $S$. How to make this change clearer?

To begin with, we avoid any hypothesis on the transformed Hamiltonian. Under this perspective, we look for that canonical transformation such that the new Hamiltonian $K$ differs from $H$ by a numerical constant $\Delta E$

$$H = K + \Delta E \tag{26}$$

The choice is possible because the energy is defined within an additive constant and, besides, we are at liberty to decide later the constraint on $K$. More importantly, the advantage of Eq. (26) resides in the correspondence with Eq. (11)

$$\frac{\partial S}{\partial t} = -\Delta E \tag{27}$$

where the energy difference $\Delta E$ replaces the original energy $E$.

Once again, we have to find the generating function of the transformation. This is calculated from the energy conservation

$$H\left(q, \frac{\partial S}{\partial q}\right) = E \tag{28}$$

where the dependence on the momenta $p$ has been replaced by the derivative of $S$ in agreement with Eq. (5). We observe that Eq. (28) is nothing but the restricted Hamilton-Jacobi equation written for the Hamilton's principal function $S$ instead of the Hamilton's characteristic function $W$ (see Eq. (12)). In other words, the restricted Hamilton-Jacobi equation is unaltered thanks to the principle of energy conservation. However, some changes are found in the formal definition of $S$. In particular, letting $H = K + \Delta E$, we find the following result

$$\frac{dS}{dt} = p_i \dot{q}_i + Q_i \dot{P}_i + \frac{\partial S}{\partial t} = p_i \dot{q}_i - \Delta E = p_i \dot{q}_i - H + K = L + K \tag{29}$$

that is analogous to Eq. (8) and would coincide with it only if $K$ was zero. An interesting feature of Eq. (29) is that the time derivative of $S$ vanishes when $\Delta E = p_i \dot{q}_i$. Finally, the Hamilton's principal function can be found as

$$S = \int p_i dq_i - \Delta E\, t = \int p_i \dot{q}_i\, dt - (H - K)t = w(t) - (H - K)t \tag{30}$$

which would be identical to Eq. (9) for $K = 0$. Note that $w(t)$ is the Hamilton's characteristic function $W$ when its natural dependences on the coordinates $q$ are viewed as functions of time. At this point, given the arbitrariness on the choice of the transformed Hamiltonian, we can safely choose $K = 0$ in agreement with the common derivation of the Hamilton-Jacobi equation. For identical reasons of free choice, the common resolution of $H(q, p) = K(P)$ can always be made after the condition $\Delta E = 0$ in Eq. (26) so that this alternative is coherently applied to Eq. (6) which, in turn, leads to the results of Eqs. (16) and (17).

The final message to convey is that the structure of the Hamilton's principal function $S$ depends on which representation of $K$ is taken. More basically, here we try to stress that the general Hamilton's principal function $S$, given in Eq. (30) for any value of the difference between $H$ and $K$, takes two specific functional dependences: one is the Hamilton's characteristic function $W$ provided that $H = K$ and the other, occurring for $K = 0$, is what is commonly called the Hamilton's principal function (and, despite Eq. (30), we will continue this tradition). Although the conclusion does not add anything new to classical mechanics, the relationship between $S$ and $W$ under an unitary vision about Hamilton's functions will reveal aspects of classical physics that hint at a connection to quantum physics. To purse the goal in a practical way, the best pedagogical example is the one-dimensional harmonic oscillator that is treated next.

# V. APPLICATION TO THE ONE-DIMENSIONAL HARMONIC OSCILLATOR

For its crucial relevance in physics, the one-dimensional harmonic oscillator represents a paradigmatic problem that serves, at the same, the purpose of illustrating the Hamilton-Jacobi technique for solving the motion of mechanical systems. The one-dimensional harmonic oscillator is also very instructive to grasp the technical differences between the two Hamilton-Jacobi methods based on the alternative choices of $K = 0$ and $K \neq 0$. Inevitably, the solution to the motion is independent from which transformed Hamiltonian is picked up and one might ask if such a freedom of choice reveals more meaning than the subsidiary use of each of the two Hamilton's functions. To highlight the points of convergence and divergence of the two methods, in this section, we first retrace some technical aspects of the calculation that are later employed to bring out features that are disregarded in more complete explanations available in the specific literature.

First of all, we recall that the conventional approach banks on the transformed Hamiltonian $K$ numerically coincident with the original Hamiltonian $H$. Under this premise, the solution of the problem is arranged in relation to the assumption $K = E = P$. In words, the energy coincides with the new conjugate momentum. Thus, the Hamilton's equation $\dot{Q} = 1$ is easily manipulated to get, after integration, $Q = t + t_0$ with $t_0$ a constant time. But $Q$ is also equal to the derivative of the Hamilton's characteristic function ($Q = \partial W / \partial P$ with $P = E$) that is known to be

$$W = \frac{E}{\omega}\left(\sqrt{\frac{U(q)}{E}\left(1 - \frac{U(q)}{E}\right)} + Arcsin\sqrt{\frac{U(q)}{E}}\right) \qquad (31)$$

where $U(q) = m\omega^2 q^2/2$. In this manner, from $Q = \partial W / \partial E$, we find

$$t + t_0 = \frac{1}{\omega} Arcsin\sqrt{\frac{U(q)}{E}} \qquad (32)$$

that, after its inversion and the use of $p = \partial W / \partial q$, leads to the recognized solution

$$q = \sqrt{\frac{2E}{m\omega^2}} \sin(\omega t + \varphi) \tag{33}$$

$$p = \sqrt{2mE} \cos(\omega t + \varphi) \tag{34}$$

with $\varphi = \omega t_0$.

If we solve the problem with the aid of $K = 0$, the calculation is slightly different. Under this circumstance, it is compulsory to adopt the Hamilton-Jacobi method based on the Hamilton's principal function $S$ that solves the so-called Hamilton-Jacobi equation without restriction. Its known solution

$$S = \frac{E}{\omega}\left(\sqrt{\frac{U(q)}{E}\left(1 - \frac{U(q)}{E}\right)} + Arcsin\sqrt{\frac{U(q)}{E}}\right) - Et \tag{35}$$

is combined with the Hamilton's equations of motion that translate into $Q = Q_0$ and $P = P_0$ with $Q_0$ and $P_0$ constants of the motion in the $(Q,P)$ phase space. Thanks to the arbitrariness of $P$, we choose again $P = P_0 = E$ and the identification of $Q_0$ with the constant time $t_0$ is mandatory. Then, the use of $Q_0 = t_0 = \partial S/\partial E$ results in the equation

$$t_0 = \frac{1}{\omega} Arcsin\sqrt{\frac{U(q)}{E}} - t \tag{36}$$

that is nothing but Eq. (32) obtained for $K \neq 0$. At this point, the procedure conforms to what shown in the previous case of $K \neq 0$ except for the employment of $S$ in place of $W$. Of course, the two Hamilton's functions are related through

$$S = W - Et \tag{37}$$

and, in the end, the two Hamilton-Jacobi methods are equivalent to each other when it comes to the solution to the motion given in terms of the old coordinate and momentum.

If the two methods of solution are very similar and lead to the same conclusion, they differ in the representative function (i.e., $S$ for $K = 0$ and $W$ for $K \neq 0$) and in terms of Hamilton's equation of motion that characterizes the new coordinate $Q$ (i.e., the fixed time $t_0$ for $K = 0$ and the variable time $t$ for $K \neq 0$). This difference has some consequences when we look for the overall time dependences of $W$ and $S$ that are rewritten as

$$w(t) = \int_{t_0}^{t} p\dot{q}d\tau = \frac{E}{\omega}\left\{\omega(t - t_0) + \frac{1}{2}\sin[2(\omega t + \varphi)] - \frac{1}{2}\sin[2(\omega t_0 + \varphi)]\right\} \quad (38)$$

$$s(t) = w(t) - Et = \frac{E}{2\omega}\{\sin[2(\omega t + \varphi)] - \sin[2(\omega t_0 + \varphi)]\} \quad (39)$$

The two functions are plotted in Fig. 1 for $t_0 = 0$ (note that the letters for the two Hamilton's functions have been changed to take into account the different functional dependences in comparison to the original dependences of $W$ and $S$). Normalization on both axes is used.

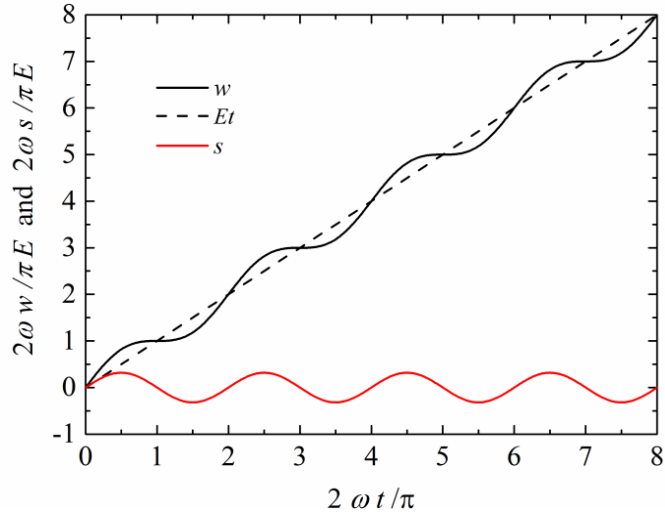

Fig. 1 Plots of $w(t)$ (black solid line) and $s(t)$ (red solid line) according to Eqs. (38) and (39). The dashed line represents the linear increase of $w(t)$. Both axes have been normalized.

It is interesting to emphasize that the Hamilton's characteristic function $w(t)$ (black solid line) oscillates around a linear increase (dashed line) and shows plateaus appearing at each odd value of the vertical axis. This second property compares well to one property that is typical of a quantum harmonic oscillator (i.e., a dependence on odd integers). On the other hand, the Hamilton's principal function oscillates in the same manner around zero without the linear increase. In other words, $w(t)$ describes an oscillator that shows time-dependent jumps specified by odd integers, whereas $s(t)$ describes an oscillator centered on its average at zero. This difference is strictly connected with the choice of $K$.

The plot of the Hamilton's characteristic function can be examined in more detail. Its linear increase as a function of the phase $\omega t$ depends on the value of the ratio between the energy and the frequency. Thus, we can calculate the energy as follows

$$E = P\omega \qquad (40)$$

and now $P$ comes to define a constant angular momentum (not to be confused with the generalized momentum defined before). It is worth noticing that Eq. (40) can already be found on the basis of the theory of canonical transformations (see for example page 379 of Ref. 1) and represents a fundamental result of the classical description of a harmonic oscillator. Intriguingly, Eq. (40) seems to mimic one fundamental rule of the quantum picture according to which the energy is proportional to the angular frequency. It is understood that such a correspondence is only qualitative or, at best, suggestive of an analogy. As a matter of fact, here, the proportionality is mediated by an unspecified angular momentum whereas in the quantum picture the proportionality is mediated by the reduced Planck constant.

Another thought-provoking aspect of Eq. (40) is its use to characterize the plateaus of $w(t)$. In Fig. 1, the plateaus appear at those times $t_n$ such that $w(t_n) = \pi P(n + 1/2)$ with $n$ indicating a generic integer number. This result is analogous to the matching condition of the Wentzel-Kramers-Brillouin (WKB) approximation (see Eq. VI.54 of Ref. 9). The obvious difference is again in the fact that, in the WKB approach, $P$ is replaced by the reduced Planck constant consequent to the application of the WKB method to Schrödinger's wave mechanics. Another point of interest stems from the constant value at zero of the Hamilton's principal function $s(t)$ when $t = t_n$ (see Fig. 1). The interest resides in the fact that a constant value for the Hamilton's principal function has a precise mechanical meaning when the wave interpretation of $S$ is recalled.[4,8,9] Indeed, the spatial derivative to the curve defined by $S = constant$ (or the gradient applied to the surface $S = constant$ in case of multi-dimensional configuration spaces) provides the direction of the motion and we do not lose the general meaning of our reasoning if we limit the analysis to the constraint of $S(q, P, t) = s(t) = 0$ occurring at $t = t_n$. We will see soon that such a constraint will be the key to the fulfillment of our ultimate goal. At present, it suffices to observe that the equation $s(t) = 0$ is surely satisfied if $w(t_n) - Et_n = 0$. Setting $Et_n = E_n T$ with $T = \pi/\omega$ the period of the oscillations in $s(t)$, we find $E_n = P\omega(n + 1/2)$ which is in close resemblance to the energy of the quantum harmonic oscillator and would coincide with it if the constant angular momentum $P$ was equal to the reduced Planck constant.

## VI. CORRESPONDENCE WITH WAVE MECHANICS OF THE FREE PARTICLE

The intriguing properties of the Hamilton's characteristic function $W$ (or its time-dependent version $w$) calculated for the harmonic oscillator are stimulating and might be the sign of a stronger correspondence with wave mechanics. Before trying to search for such a correspondence, let us summarize the basic Hamilton-Jacobi machinery that might be useful to make this stronger correspondence emerge. We start with the simplest example of a free particle. The treatment of the harmonic oscillator is instead postponed until next Section.

The free particle is described by the Hamiltonian $H = p^2/2m$ that equals the total energy $E$. The cyclic condition means that the linear momentum $p$ is a constant and the Hamilton's equations for the choice $K = 0$ imply that the new conjugate variables $Q$ and $P$ are constants too. Given this transformed Hamiltonian, the further choice $P = E$ means that the transformed coordinate is a fixed time, $Q = t_0$. On the other hand, based on the other choice $K = H = E$, the main difference is in the meaning of the coordinate $Q$ that becomes the time variable $t$ when we keep the identification of the transformed momentum $P$ with the energy $E$. Then, based on $p = \partial W/\partial q$, we find the Hamilton's functions

$$W = \pm q\sqrt{2mE} = qp \tag{41}$$

$$S = \pm q\sqrt{2mE} - Et \tag{42}$$

where $p = \pm\sqrt{2mE}$ has been used in Eq. (41). The introduction of $Q = \partial S/\partial E$ corresponds to the following fixed time

$$t_0 = \frac{mq}{p} - t \tag{43}$$

and the inversion of Eq. (43) gives

$$q = \frac{p}{m}(t + t_0) \tag{44}$$

that is equivalent to taking the initial coordinate $q_0 = pt_0/m$ at the time $t = 0$.

The calculation shown above is very simple and can be found in textbooks. The true element of novelty is represented by what follows. Indeed, at this point, the use of Eq. (44) helps us to write the Hamilton's functions in terms of the whole set of old and new variables taken for the choice of $K = 0$, namely $q$, $p$, $t_0$ and $E$. After some algebraic manipulation (see Appendix A), we find the following expressions for the Hamilton's functions

$$\Psi_W = \frac{qp}{2} + QP\left(1 + \frac{t}{t_0}\right) \tag{45}$$

$$\Psi_S = \frac{qp}{2} + QP \tag{46}$$

In Eqs. (45) and (46), the letter $\Psi$ symbolizes the different functional dependences appearing in $W$ and $S$ after the algebraic operations (modified Hamilton's functions). In addition, the use of $Q$ and $P$ instead of $t_0$ and $E$ is preferred to underline the dependence on the new conjugate variables defined by the choice of $K = 0$ (the symbol of $t_0$ is purposefully left in the brackets of Eq. (45) to visualize immediately the dimensionless ratio $t/t_0$). At first sight, the use of $Q = t_0$ in the product $QP$ of $\Psi_W$ might appear contradictory. As a matter of fact, the Hamilton's characteristic function $W$ or its equivalent $\Psi_W$ are connected to the choice of $K \neq 0$, which calls for $Q = t$ as natural variable. However, here we follow a different route and we ask what the expressions of both Hamilton's functions are when we simultaneously use the old and new variables that characterize the canonical transformation of Eq. (3) for the choice $K = 0$. Clearly, the advantage resides in the constant value of $t_0$ and $E$ (that is, $Q$ and $P$) so that the true time-dependent variables are only $q$ and $p$ for both modified Hamilton's functions $\Psi_W$ and $\Psi_S$.

Now, we notice that the product $QP$ in Eqs. (45) and (46) gives the area of the rectangle identified by the specific constant values of $Q$ and $P$ in the corresponding phase space. This means that

$$QP = \int dQ dP = J_1 \qquad (47)$$

where $J_1$ is the first Poincaré invariant.[1] In the end, the expressions of the modified Hamilton's functions in dependence of the old variables and $J_1$ are

$$\Psi_W = \frac{qp}{2} + J_1\left(1 + \frac{t}{t_0}\right) \qquad (48)$$

$$\Psi_S = \frac{qp}{2} + J_1 \qquad (49)$$

Next, we focus our attention on those values of the linear momentum $p$ such that $\Psi_S$ is a constant and, without losing generality, we can set this constant equal to zero so that the linear momentum is inversely proportional to the coordinate $q$

$$p = -\frac{2J_1}{q} \qquad (50)$$

The interest in a stationary Hamilton's principal function comes from Eq. (8) and from the wave interpretation of the Hamilton's functions.[4,8,9] In particular, the surfaces in the configuration space created by the condition $S = constant$ are important because the unit vectors normal to the surfaces define the direction of the motion. Then, we choose $S = \Psi_S = 0$ and we can ask the question of how the Hamilton's characteristic function $W$ changes if the principal function $S$ is zero. This is the central question that stands behind of what follows. Under this constraint, $p$ satisfies Eq. (50) and the fundamental property $p = \partial W/\partial q$ yields the following equation for the Hamilton's characteristic function $W$

$$\frac{\partial^2 W}{\partial q^2} = \frac{\partial p}{\partial q} = \frac{2QP}{q^2} = -\frac{W}{4J_1^2}p^2 \qquad (51)$$

where in the last passage we have used $2QP = -qp = -W$ on the basis of Eq. (41) conditioned to $S = \Psi_S = 0$. With the help of $p^2 = 2mE$, Eq. (51) becomes

$$-\frac{2J_1^2}{m}\frac{\partial^2 W}{\partial q^2} = EW \tag{52}$$

This equation looks similar to the time-independent Schrödinger's equation of the free particle. Indeed, the Hamilton's characteristic function $W$ takes the role of the Schrödinger's wave function. Such a correspondence is somewhat expected by virtue of physical and historical reasons (the connection with Hamilton's characteristic function appears at the very beginning of wave mechanics[10]). Despite this, the Poincaré invariant $J_1$ appears instead of $\hbar/2$. Although $J_1$ is constant, it is obvious that the appearance of $\hbar$ is peculiar to quantum mechanics only. On the other hand, here we look for a formal correspondence and it is not really crucial for the correspondence to work if a constant is replaced by another constant.

## VII. CORRESPONDENCE WITH WAVE MECHANICS OF THE QUANTUM HARMONIC OSCILLATOR

The connections of the Hamilton's characteristic function $W$ (or its time-dependent version $w$ plotted in Fig. 1) with Schrödinger's wave mechanics are of inspiration to search for a stronger correspondence along the lines of investigation described in the previous section. To this end, we rewrite both Hamilton's functions $W$ and $S$ with the help of Eqs. (32)-(34), then Eqs. (31) and (35) become respectively (see Appendix B)

$$\Psi_W = \frac{qp}{2} + E(t + t_0) \tag{53}$$

$$\Psi_S = \frac{qp}{2} + Et_0 \tag{54}$$

where the letter $\Psi$ symbolizes the modified Hamilton's function having different functional dependences in comparison with $W$ and $S$. Now, we know that if we were to solve the Hamilton-

Jacobi equation by using the canonical transformation for the transformed Hamiltonian $K = 0$, then the new conjugate coordinate $Q$ and the new conjugate momentum $P$ would be respectively the time $t_0$ and the energy $E$. Then, Eqs. (53) and (54) reduce to

$$\Psi_W = \frac{qp}{2} + QP\left(1 + \frac{t}{t_0}\right) \tag{55}$$

$$\Psi_S = \frac{qp}{2} + QP \tag{56}$$

where both, $Q$ and $P$, are constants of the motion. The result summarized in Eqs. (55) and (56) is striking. The Hamilton's functions of the harmonic oscillator, when written in dependence on the old and new variables relative to the choice $K = 0$ (modified Hamilton's functions), are identical to those of the free particle reported in Eqs. (45) and (46). This means that there exists a general equivalence between the free propagation and the harmonic oscillator provided that the equivalence is limited to the condition of constant Hamilton's principal function (that is, wave fronts characterized by $S = constant$). This equivalence is striking because the harmonic oscillator is characterized by a force that is totally absent for the free particle. On the other hand, we are accustomed to hearing that free-particle behavior is intertwined with oscillatory (or wavy) behavior (one example among many is the Young's experiment with single photons or electrons). Here, the equivalence established between Eqs. (45)-(46) and Eqs. (55)-(56) gives the ground for some insight into the connection between free propagation and oscillators.

The finding has, indeed, an immediate consequence. We can repeat the same procedure leading to Eq. (51) except for the substitution of the linear momentum according to $p^2 = 2m[E - U(q)]$ and, in the end, we have

$$-\frac{2J_1^2}{m}\frac{\partial^2 W}{\partial q^2} = [E - U(q)]W \tag{57}$$

that establishes the sought correspondence with wave mechanics. Once again, the fact that we have $2J_1^2$ in place of $\hbar^2/2$ is not disturbing in that we are dealing with a classical problem. In both cases, the representative parameter, $J_1$ or $\hbar$, acts as a constant regardless of its numerical value.

Despite the result of Eq. (57), the presence of the factor 2 is problematic. It can be suppressed though! It disappears when we consider a two-dimensional harmonic oscillator described in general by the following Hamilton-Jacobi equation

$$\frac{1}{2m}\left[\left(\frac{\partial W}{\partial x}\right)^2 + m^2\omega_x^2 x^2 + \left(\frac{\partial W}{\partial y}\right)^2 + m^2\omega_y^2 y^2\right] = E \tag{58}$$

The conventional solution to Eq. (58) is based on the separation of variables so that two independent Hamilton-Jacobi problems are defined after the energy splitting $E = E_x + E_y$

$$\frac{1}{2m}\left[\left(\frac{\partial W}{\partial x}\right)^2 + m^2\omega_x^2 x^2\right] = E_x \tag{59}$$

$$\frac{1}{2m}\left[\left(\frac{\partial W}{\partial y}\right)^2 + m^2\omega_y^2 y^2\right] = E_y \tag{60}$$

The solutions to Eqs. (59) and (60) are identical to Eq. (31) and they are summed to give

$$W = W_1(x, E_x) + W_2(y, E_y) \tag{61}$$

Considering first Eq. (59), we have

$$W_1(x, E_x) = \frac{E_x}{\omega_x}\left(\sqrt{\frac{U(x)}{E_x}\left(1 - \frac{U(x)}{E_x}\right)} + Arcsin\sqrt{\frac{U(x)}{E_x}}\right) \tag{62}$$

The component $W_2(y, E_y)$, which appears as an additive constant in the solution of Eq. (59), is found by means of Eq. (60). In the end, the whole solution is then

$$W = W_1(x, E_x) + W_2(y, E - E_x) \tag{63}$$

If we elaborate more Eq. (63) along the lines described before, we find that the modified Hamilton's principal function is

$$\Psi_S = \frac{xp_x + yp_y}{2} + Et_0 \tag{64}$$

When the oscillator is perfectly symmetric, we can set $\omega_x = \omega_y = \omega$ and $E_x = E_x = E/2$. Based on the symmetry between the two oscillators, we can define $x = y = q$ and $p_x = p_y = p$ and then Eq. (64) becomes

$$\Psi_S = qp + QP \tag{65}$$

where we have set $t_0 = Q$ and $E = P$. At this point, we find for $W_1$

$$-\frac{J_1^2}{2m}\frac{\partial^2 W_1}{\partial x^2} = [E_x - U(x)]W_1 \qquad (66)$$

and a similar equation for $W_2$. This equation is the exact copy of the Schrödinger's equation apart from the obvious replacement of $J_1$ with $\hbar$.

### VIII. GENERAL CORRESPONDENCE WITH WAVE MECHANICS

The examples examined in the previous sections introduce the question of whether the correspondence with wave mechanics is more general than the cases of a free particle and a harmonic oscillator. Reviewing the procedure that supports the correspondence, the key role is played by the equation

$$qp = constant \qquad (67)$$

This equation stems from the stationary condition on the Hamilton's principal function $S$ and, in turn, involves constant values for the transformed variables $Q$ and $P$ ($K = 0$ is always associated with $S$). It is then worth noticing that the condition in Eq. (67) is obeyed by the seminal solutions to Schrödinger's equation. For instance, the solution to the quantum harmonic oscillator is notoriously given by

$$\psi = \psi_0 e^{-\frac{x^2}{2}} H_n(x) \qquad (68)$$

with $x = q\sqrt{m\omega/\hbar}$. Then, within the language of the Hamilton-Jacobi approach, we can take the new conjugate variables as follows: $P = E/\omega = \hbar$ and $Q = \omega t_0$ with $t_0$ a given time instant. The new conjugate variables are actually constant because they define the Hamilton's equations for the transformed Hamiltonian $K = 0$. For this reason, after we constrain $S$ to zero (or $\Psi_S = 0$ in Eq. (56)), we expect that the old variables $q$ and $p$ obey the equation of the classical harmonic oscillator (i.e., $qp = -2QP$). Let us verify it for the Schrödinger's wave function of Eq. (68). On the basis of a proportionality between $\psi$ and the Hamilton's characteristic function $W$ suggested in the correspondence established in Eqs. (52), (57) and (66) (that is, $\psi = aW$), we find

$$qp = q\frac{\partial W}{\partial q} = \frac{q}{a}\frac{\partial \psi}{\partial q} = \frac{q}{a}\frac{\partial \psi}{\partial x}\frac{\partial x}{\partial q} = \frac{x}{a}\frac{\partial \psi}{\partial x} \tag{69}$$

and

$$QP = P\frac{\partial W}{\partial P} = \frac{P}{a}\frac{\partial \psi}{\partial x}\frac{\partial x}{\partial P} = \frac{\hbar}{a}\frac{\partial \psi}{\partial x}\frac{\partial x}{\partial \hbar} = -\frac{x}{2a}\frac{\partial \psi}{\partial x} \tag{70}$$

It is now easy to combine Eqs. (69) and (70) to obtain $qp = -2QP$. This result suggests the idea that the Schrödinger's wave function acts as it was the generating function of the canonical transformation between the old and new variables.

The very same conclusion of $qp = constant$ can be reached if we consider the solution to the radial Schrödinger's equation of the hydrogen atom. In this case

$$\psi_R = R_0 e^{-\frac{\rho}{2}} \rho^l L_{n+l}^{2l+1}(\rho) \tag{71}$$

with $R_0$ a normalization constant, $\rho = \alpha_n r$ and $\alpha_n = 2me^2/4\pi\varepsilon_0\hbar^2 n$. Once again, considering the Hamilton-Jacobi variables $q = r$, $P = E/\omega = \hbar$ and $Q = \omega t_0$ taken for the Hamilton's principal function $S$, the condition of a proportionality between $W$ and $\psi_R$ implies that

$$qp = q\frac{\partial W}{\partial q} = \frac{r}{a}\frac{\partial \psi_R}{\partial r} = \frac{r}{a}\frac{\partial \psi_R}{\partial \rho}\frac{\partial \rho}{\partial r} = \frac{\rho}{a}\frac{\partial \psi_R}{\partial \rho} \tag{72}$$

and

$$QP = P\frac{\partial W}{\partial P} = \frac{P}{a}\frac{\partial \psi_R}{\partial \rho}\frac{\partial \rho}{\partial P} = \frac{\hbar}{a}\frac{\partial \psi_R}{\partial \rho}\frac{\partial \rho}{\partial \hbar} = -2\frac{\rho}{a}\frac{\partial \psi_R}{\partial \rho} \tag{73}$$

so that we find $qp = -QP/2 = constant$.

The recurrence of Eq. (67) in fundamental results of Schrödinger's wave mechanics stimulates the idea of a generalization. Indeed, we can calculate the time derivative of Eq. (67) as follows

$$\dot{q}p + q\dot{p} = 0 \tag{74}$$

and, introducing the definition of the conjugate momentum through the Euler-Lagrange approach of classical mechanics, we find

$$\dot{q}\frac{\partial L}{\partial \dot{q}} + q\frac{\partial L}{\partial q} = 0 \qquad (75)$$

Eq. (75) can be made explicit by means of the Lagrangian $L$ when the potential energy $U$ has a power-law dependence ($U = kq^n$)

$$L = \frac{1}{2}m\dot{q}^2 - kq^n \qquad (76)$$

with $k$ a constant whose sign is negative or positive for attractive or repulsive forces, respectively. After the insertion of Eq. (76) into Eq. (75), we obtain

$$2T - nU = 0 \qquad (77)$$

which is the exact copy of the statistical equation known by the name of the virial theorem.[1] In other terms, Eq. (77) establishes the exact relationship between kinetic and potential energies whenever the product $qp$ is constant and, vice versa, if Eq. (77) holds true then Eq. (67) holds true also. Note that, here, the key remark is on the exactness of Eq. (77). By contrast, if we release the constraint on the product $qp$, Eq. (77) can only be found in statistical sense. Under this last circumstance, time averages $\langle T \rangle$ and $\langle U \rangle$ appear in the same equation that is popularly recognized as the virial theorem. This finding bears the additional consequence that, when the virial theorem applies, then the most probable trajectory is that one associated with Eq. (67). However, none of this statistical extension matters in the current context. As long as Eq. (67) is satisfied, the result of Eq. (77) emphasizes that kinetic and potential energies appear with their exact value and no averages are involved. At this point, we have a clear connection between Eq. (77), Eq. (67) and the condition of $S = 0$ (or $S = constant$ that implies a change in the origin of the time axis).

Next, let us see if we are able to generalize the connection with wave mechanics. Actually, this further step is rather easy. It suffices to focus our attention on Eq. (77) according to which the Hamiltonian becomes $H = E = (1 + 2/n)T$. This Hamiltonian is cyclic and related to the Hamiltonian of the free particle treated before. Fundamental examples are the harmonic oscillator ($n = 2$) and Kepler-type problems (gravitational or electric forces are defined by $n = -1$). The case of the harmonic oscillator resulting from the perfect identity

$$T = U \tag{78}$$

has been properly considered in the previous section. For Kepler-type problems, Eq. (77) is

$$T = -U/2 \tag{79}$$

that yields

$$H = E = -T = -\frac{p^2}{2m} \tag{80}$$

Eq. (80) is well known in the classical treatment of the Kepler problem and suggests that the constraint $qp = constant$ produces a cyclic Hamiltonian that is in clear analogy with the Hamiltonian of the free particle having imaginary linear momentum, or $p_{free} = \pm ip$. Note that Eq. (80) might be considered as a justification for the sudden appearance of the imaginary unit in the definition of the linear momentum. More importantly for our argument, the cyclic condition determines a constant $p$ and the Hamilton's functions become

$$W = pq \tag{81}$$

$$S = pq - Et \tag{82}$$

Then, Eq. (43) or (44) can be used and we get

$$\Psi_W = \frac{qp}{2} + QP\left(1 + \frac{t}{t_0}\right) \tag{83}$$

$$\Psi_S = \frac{3}{2}pq + QP \tag{84}$$

where $Q = t_0$ and $P = E$. The final conclusion is that the requirement $S = \Psi_S = 0$ leads to Eq. (67) and we can repeat the same procedure shown before to obtain a Schrödinger-type equation for the Hamilton's characteristic function.

**IX. CONCLUSIONS**

In this work, a revision of the Hamilton-Jacobi theory is made to understand the changes caused by more general representations of the transformed Hamiltonian $K$. For mechanical problems characterized by the time-dependent Hamiltonian $H$ an equally time-dependent transformed Hamiltonian is shown to provide the same Hamilton-Jacobi equation obtained for the typical choice of $K = 0$. Furthermore, for conservative mechanical systems with constant $H$, it is shown that the Hamilton's principal function can be generally defined on the basis of the difference between $K$ and $H$. This amounts to say that the generalized principal function can take specific dependences. In particular, it coincides with the characteristic function $W$ when $K = H$, or with the ordinary principal function $S$ when $K = 0$. The intimate relationship between $W$ and $S$ is examined in a new light for the fundamental example of a one-dimensional classical harmonic oscillator. The relationship reveals that there are a few fundamental points of contact with features that are peculiar to the quantum version of the oscillator. These features emerge from the constraint of $S = 0$. The constraint is related to the dynamical information contained in the spatial derivative of $S$ and, for this reason, we have analyzed this constraint in more detail. When the original and transformed conjugate variables associated with the Hamilton's principal function are introduced to describe both Hamilton's functions, it appears that the constraint $S = 0$ induces a Schrödinger's equation where the Hamilton's characteristic function $W$ plays the role of the Schrödinger's wave function, whereas the first integral invariant of Poincaré takes the place of the reduced Planck constant. This happens not only for the harmonic oscillator and seems to be a general rule connected to the equation $qp = constant$ where $q$ and $p$ are the conjugate variables of the one-dimensional mechanical problem specified by the original Hamiltonian $H$. As a matter of fact, the equation $qp = constant$ is demonstrated to be one way to state the virial theorem in which averages of the kinetic and potential energies are replaced by their exact values. For this reason, the conclusions reached for the harmonic oscillator can be extended to those mechanical problems that obey the virial theorem.

   The findings contain in this work make the quantum-classical correspondence between Schrödinger's wave mechanics and Hamilton-Jacobi theory much stronger than before and suggest that the Schrödinger's wave function might be interpreted as the generating function of the canonical transformation that connects variables in the real phase space to the transformed variables that have $\hbar$ as the constant angular momentum $P$ associated with the choice of an identically zero transformed Hamiltonian. Additional work is planned to extend the ideas laid out here to mechanical problems of higher spatial dimensions.

**Appendix A**

From Eqs. (41) and (44)

$$W = \frac{qp}{2} + \frac{p^2}{2m}(t + t_0) \tag{A1}$$

and, for the free particle, we obtain

$$W = \frac{qp}{2} + \frac{p^2}{2m}(t + t_0) = \frac{qp}{2} + Et_0\left(1 + \frac{t}{t_0}\right) \tag{A2}$$

The substitution of Eq. (A2) in Eq. (42) yields

$$S = \frac{qp}{2} + Et_0 = \frac{qp}{2} + QP \tag{A3}$$

where $Q = t_0$ and $P = E$ is the choice made for the new conjugate variables corresponding to the principal Hamilton's function. The two Eqs. (A2) and (A3) are formally incorrect because the Hamilton's functions $W$ and $S$ imply mathematical dependences on $q$ and $P$ only. For this reason, the correct expressions should be given in terms of functions that have the same numerical values of $W$ and $S$ with different functional dependences (modified Hamilton's functions). The correction correspond to writing Eqs. (45) and (46) given in the main text.

**Appendix B**

The Hamilton's characteristic function of the harmonic oscillator can be written in terms of $q$ and $p$ variables. To do so, the potential energy is modified with the help of Eq. (33)

$$U(q) = \frac{1}{2}m\omega^2 q^2 = E\,sin^2(\omega t + \varphi) \tag{B1}$$

and its substitution in the first term of Eq. (31) provides

$$\frac{E}{\omega}\sqrt{\frac{U(q)}{E}\left(1-\frac{U(q)}{E}\right)} = \frac{E}{\omega}sin(\omega t + \varphi)cos(\omega t + \varphi) \tag{B2}$$

However, thanks to Eqs. (33) and (34) solutions to the motion of the harmonic oscillator, Eq. (B2) becomes

$$\frac{E}{\omega}\sqrt{\frac{U(q)}{E}\left(1-\frac{U(q)}{E}\right)} = \frac{qp}{2} \tag{B3}$$

and therefore the Hamilton's characteristic functions takes the following appearance

$$W = \frac{qp}{2} + E(t+t_0) = \frac{qp}{2} + Et_0\left(1+\frac{t}{t_0}\right) \tag{B4}$$

where Eq. (32) or (36) has been used. The result in Eq. (B4) is reported as Eq. (53) in the main text after the symbolic change, from $W$ to $\Psi_W$, needed to emphasize the change in the functional dependences of the Hamilton's characteristic function. Similarly, the corresponding expression of the Hamilton's principal function

$$S = W - Et = \frac{qp}{2} + Et_0 \tag{B5}$$

is reported as $\Psi_S$ in Eq. (54) of the main text to mean the different functional structure when the Hamilton's principal function is viewed in dependence on $q$ and $p$.

It is important to note that Eqs. (B4) and (B5) obtained for the harmonic oscillator are identical to those of the free particle (Eqs. (A2) and (A3)).

---

*michele.marrocco@enea.it; michele.marrocco@uniroma1.it